\documentclass{llncs}
\usepackage{graphicx}
\usepackage{times}
\usepackage{url}
\usepackage{amsmath}
\usepackage{hyperref}

\def\hexdigit#1{\ifnum#1<10 \number#1\else
\ifnum#1=10 A\else\ifnum#1=11 B\else\ifnum#1=12 C\else
\ifnum#1=13 D\else\ifnum#1=14 E\else\ifnum#1=15 F\fi
\fi\fi\fi\fi\fi\fi}
\font\tenmsam=msam10

\font\sevenmsam=msam7
\font\fivemsam=msam5
\newfam\msafam
\textfont\msafam=\tenmsam
\scriptfont\msafam=\sevenmsam
\scriptscriptfont\msafam=\fivemsam

\font\tenmsbm=msbm10

\font\sevenmsbm=msbm7
\font\fivemsbm=msbm5
\newfam\msbfam
\textfont\msbfam=\tenmsbm
\scriptfont\msbfam=\sevenmsbm
\scriptscriptfont\msbfam=\fivemsbm
\def\mathbb{\fam=\msbfam\tenmsbm}

\mathchardef\N="0\hexdigit\msbfam4E
\mathchardef\R="0\hexdigit\msbfam52
\mathchardef\Z="0\hexdigit\msbfam5A
\mathchardef\F="0\hexdigit\msbfam46
\mathchardef\le="3\hexdigit\msafam36
\mathchardef\ge="3\hexdigit\msafam3E

\def   \mod    #1{\;({\rm mod}\;#1)}

\def   \beas   {\begin{eqnarray*}}
\def   \eeas   {\end{eqnarray*}}

\begin{document}

\title{Cloud Data Auditing Using Proofs of Retrievability\thanks{This is an author's version of the chapter published in Guide to Security Assurance for Cloud Computing (Springer International Publishing Switzerland 2015). The final publication is available at \href{https://link.springer.com/chapter/10.1007\%2F978-3-319-25988-8_11}{link.springer.com}.}}
\author{Binanda Sengupta \and Sushmita Ruj}
\index{Sengupta, Binanda}
\index{Ruj, Sushmita}
\institute{Indian Statistical Institute\\Kolkata, India\\
{\tt \{binanda\_r,sush\}@isical.ac.in}}

\maketitle

\pagestyle{plain}

\begin{abstract}
Cloud servers offer data outsourcing facility to their clients. A client outsources her data
without having any copy at her end. Therefore, she needs a guarantee that her data are not
modified by the server which may be malicious. Data auditing is performed on the outsourced
data to resolve this issue. Moreover, the client may want all her data to be stored untampered.
In this chapter, we describe proofs of retrievability (POR) that convince the client about
the integrity of all her data.

\end{abstract}

\section{Introduction}

Proper storage and maintenance of data has been an important research problem
in the field of computer science. With the advent of cloud computing, cloud service
providers (CSP) offer various facilities to their clients. For example, clients
can outsource their computational workload to the cloud server, or clients having
limited storage capacity can store huge amount of data in the server. Several
storage service providers (SSP) provide this type of storage outsourcing
facility to their clients. Amazon Simple Storage Service (S3)~\cite{AmazonS3},
Google Drive~\cite{GDrive}, Dropbox~\cite{Dropbox} are a few of them to mention.
These storage service providers (we will use SSP and server interchangeably in this chapter)
store clients' data in lieu of monetary benefits.
The nature of the outsourced data may be \textit{static} (never modified once they are
uploaded) or \textit{dynamic} (clients can modify them).
On the other hand, the servers can be
malicious to maximize the benefits with a constant amount of storage at their end.
We consider a situation as follows. Suppose, a server has 100 GB (say) of storage capability.
Now, there are two clients requesting for 100 GB of storage each to the server. The server
may store 50 GB data of each client and claim that it has stored 200 GB data. When any client
wants to download her data from the server, she gets only half of her data. Therefore, the
client needs a guarantee that her data have been stored intact by the server. In case of
any modification or deletion of data, the server needs to compensate the client appropriately.
Hence, the only question remains is that how to guarantee the integrity of the client's data.
To address the problem of checking the integrity of data, auditing comes into play.
Audits may be performed as often as asked by the client.
If a server can pass an audit, then the client is convinced that her file (or some part of it)
is stored untampered. However, a server can pass an audit without storing the file properly,
but the probability of such an event is ``very small'' (a negligible function in the security
parameter as defined in Section~\ref{notation}).

One naive way of auditing is as follows. A client does some preprocessing on her data before
uploading them to the cloud server. This preprocessing phase includes attaching some cryptographic
authenticators (or tags) corresponding to segments (or blocks) of the data file. In the following section,
we discuss about some cryptographic tools based on which these authenticators are designed.
The idea of using these authenticators is to prevent the server from modifying the file and
still passing an audit with high probability. Now, the client uploads the processed file
(data file along with the authenticators) to the storage server. During an audit, the
client downloads the whole file along with the tags to her end, and verify the authenticity
of each of the blocks. The server passes an audit if and only if each block-authenticator pair
is valid.

The naive idea mentioned above suffers from severe drawbacks. Every time a client asks for an audit,
she has to download all her data from the server which incurs a high communication bandwidth.
To overcome this issue, researchers have come up with what is called \textit{proofs of storage}.
As before, the client computes an authenticator for each block of her data (or file), and uploads the
file along with the authenticators. During an audit protocol, the client samples a predefined number of
random block-indices and sends them to the server (challenge). The server does some computations
over the challenge, stored data and authenticators, and sends a response to the client
who verifies the integrity of her data based on this response. This is an example
of \textit{provable data possession} (PDP) introduced by Ateniese et al.~\cite{Ateniese_CCS}.
This work is followed by other provable data possession schemes
\cite{Ateniese_SCOM,Curtmola_ICDCS,Erway_CCS,Ateniese_ACM}.
Though these schemes guarantee the integrity of \textit{almost all} the blocks of the data file,
PDP cannot convince the client about the integrity of \textit{all} the blocks.
The outsourced data may contain some sensitive accounting information which the client
do not want to lose. On the other hand, losing the compression table of a compressed file
makes the whole file unavailable. In such circumstances, the client wants a stronger
notion than PDP which would guarantee that the entire file has been stored properly and
the client can \textit{retrieve} her file at any point of time.

The first paper
introducing proofs of retrievability (POR) for static data is by Juels and Kaliski~\cite{JK_CCS}
(a similar idea was given for sublinear authenticators by Naor and Rothblum~\cite{NR_JACM}).
They introduce erasure coding (see Section~\ref{erasure_code}) to the proofs of storage.
The underlying idea is to encode the original file with some erasure code, compute authenticators
for the blocks of the encoded file, and then upload the file along with the authenticators to
the data server. With this technique, the server has to delete or modify a considerable
number of blocks to actually delete or modify a data block. Thus, the probability that the server
passes an audit, given some data blocks are actually deleted or modified, is ``very small''.
This technique ensures that all the blocks of the file are correctly stored at
the server's end. This notion is formalized by defining an extractor algorithm which can extract,
with high probability, the original file after interacting with a server which passes an audit with
some non-negligible probability. We review some of the POR schemes in this chapter.

The organization of this chapter is as follows. In Section~\ref{prelim}, we describe some notations and tools
which will be used in later sections. Section~\ref{por_static} discusses some POR schemes for static data.
In Section~\ref{por_dynamic}, we describe some POR schemes for dynamic data. We conclude the chapter in
Section~\ref{conclusion}.

\section{Preliminaries}\label{prelim}
In this section, we briefly discuss about some backgrounds needed for understanding
the following sections. The detailed discussions can be found
in~\cite{Stinson,KatzLindell,GoldreichI,GoldreichII,Sipser,Arora,GoldreichIII,MWSloane77}.

\subsection{Notation}\label{notation}
We take $\lambda$ as the security parameter.
An algorithm $\mathcal{A}(1^\lambda)$ is called a probabilistic polynomial-time
algorithm when its running time is polynomial in $\lambda$ and its output $y$
is a random variable which depends on the internal coin tosses of $\mathcal{A}$.
We write $y\leftarrow\mathcal{A}(\cdot)$ or $y\leftarrow\mathcal{A}(\cdot,\ldots,\cdot)$
depending upon whether $\mathcal{A}$ takes one input or more inputs, respectively.
Moreover, if $\mathcal{A}$ is given access to an oracle $\mathcal{O}$, we
write $y\leftarrow\mathcal{A}^{\mathcal{O}}(\cdot,\ldots,\cdot)$. In this
case, the Turing machine $\mathcal{A}$ has an additional query tape
where $\mathcal{A}$ places its query $x$ and calls another Turing machine
$\mathcal{O}$. Then, $\mathcal{O}$ is invoked with the input $x$, and
the output is written on the same query tape~\cite{Sipser,Arora}.
An element $a$ chosen uniformly at random from a set $S$
is denoted as $a\xleftarrow{R}S$.
A function $f:\N\rightarrow\R$ is called negligible in $\lambda$ if for
all positive integers $c$ and for all sufficiently large $\lambda$, we have
$f(\lambda)<\frac{1}{\lambda^c}$. We call a problem ``hard'' to denote
that no polynomial-time algorithms exist for solving the problem.

\subsection{Message Authentication Code (MAC)}
Let $f: \mathcal{K}\times\mathcal{M}\rightarrow\mathcal{D}$ be a function,
where the $\mathcal{K}$ is the key space, $\mathcal{M}$ is the message space
and $|\mathcal{D}|\ll|\mathcal{M}|$. In other words, $f$ takes as inputs
a secret key $k\in\mathcal{K}$ and a message $m\in\mathcal{M}$,
and it outputs $d\in\mathcal{D}$. The piece of information $d$ is a message
authentication code (MAC)~\cite{Stinson} if the following properties are satisfied.
\begin{enumerate}
 \item Given $m$ and $d$, it is hard to find another $m'\not =m$ such that
       $f_k(m)=f_k(m')$.
 \item The value of $f_k(m)$ should be uniformly distributed in the set $\mathcal{D}$.
 \item The value of $f_k(m)$ should depend on every bit of the message $m$ equally.
\end{enumerate}
Message authentication codes are used as a digest to authenticate the message.
MACs are defined in symmetric setting, that is, the sender and the receiver
need to share a secret key. In the generation phase, the sender calculates the
MAC for the message using the secret key and sends the message along with the MAC.
In the verification phase the receiver verifies, using the same key, whether the
MAC is computed on the given message using the same secret key.
Due to the first property mentioned above, it is hard to
modify a message $m$ keeping the value of $f_k(m)$ unchanged.

Message authentication codes are used hugely for authentication purposes. There are several
constructions for MACs. Some of the constructions are based on pseudorandom
functions~\cite{KatzLindell,GoldreichI,Luby_PRP} (e.g.,~XOR MAC~\cite{XORMAC_CR},~CMAC~\cite{CMAC}),
and some of them are based on cryptographic hash functions (e.g.,~HMAC~\cite{HMAC}).

\subsection{Bilinear Maps}\label{blmap}
Let $G_1,G_2$ and $G_T$ be multiplicative cyclic groups of prime order $p$.
Let $g_1$ and $g_2$ be generators of the groups $G_1$ and $G_2$, respectively.
A bilinear map~\cite{KM_CC,Lynn_Thesis,Galbraith_DAM} is a function $e: G_1\times G_2\rightarrow G_T$ such that:

1. for all $u\in G_1, v\in G_2, a,b\in\Z_p$, we have $e(u^a,v^b)=e(u,v)^{ab}$
(bilinear property),

2. $e$ is non-degenerate, that is, $e(g_1,g_2)\not = 1$.

\noindent
Furthermore, properties 1 and 2 imply that

3. for all $u_1,u_2\in G_1, v\in G_2$, we have $e(u_1\cdot u_2,v)=e(u_1,v)\cdot e(u_2,v)$.

\noindent
If $G_1=G_2=G$, the bilinear map is symmetric; otherwise, asymmetric.
Unless otherwise mentioned, we consider bilinear maps which
are symmetric and efficiently computable. Let BLSetup$(1^\lambda)$ be
an algorithm which outputs $(p,g,G,G_T,e)$, the parameters of a
bilinear map, where $g$ is a generator of $G$.

\subsection{Digital Signature}\label{dig_sig}
Diffie and Hellman introduce the public-key cryptography and the notion
of digital signatures in their seminal paper ``New Directions in
Cryptography''~\cite{DH_ITIT}. Rivest, Shamir and Adleman~\cite{RSA_CACM}
propose the first digital signature scheme based on the RSA assumption.
Several signature schemes are available in the literature.
Several signature schemes are found in the literature
\cite{Lamport_79,Rabin_79,Merkle_CR,FiatShamir_CR,ElGamal_CR,Schnorr_JOC,DSA,ECDSA_IJIS,BLS_JOC,CL_CR,CHP_JOC,ChCh_PKC,Hess_SAC,BBS_CR,CYH_ACNS,BGLS_ECR}.

We define a digital signature scheme as proposed by Goldwasser
et al.~\cite{GMR_ACM}. A digital signature scheme
consists of the following polynomial-time algorithms:
a key generation algorithm KeyGen, a signing algorithm Sign
and a verification algorithm Verify. KeyGen takes as input the security parameter
$\lambda$ and outputs a pair of keys $(pk,sk)$, where $sk$ is the
secret key and $pk$ is the corresponding public key. Algorithm
Sign takes a message $m$ from the message space $\mathcal{M}$
and the secret key $sk$ as input
and outputs a signature $\sigma$.
Algorithm Verify takes as input the public key $pk$, a message $m$
and a signature $\sigma$, and outputs \texttt{accept} or \texttt{reject}
depending upon whether the signature is valid or not.
Any of these algorithms can be probabilistic in nature.
A digital signature scheme has the following properties.
\begin{enumerate}
 \item \textit{Correctness}: Algorithm Verify always accepts a signature generated by an honest signer, that is,
	 \beas
	  \Pr[\text{Verify}(pk,m,\text{Sign}(sk,m))=\texttt{accept}]=1.
	 \eeas
 \item \textit{Security}: Let Sign$_{sk}(\cdot)$ be the signing oracle and $\mathcal{A}$
	 be any probabilistic polynomial-time adversary with an oracle access to Sign$_{sk}(\cdot)$.
	 The adversary $\mathcal{A}$ makes polynomial
	 number of sign queries to Sign$_{sk}(\cdot)$ for different messages and gets back the signatures
	 on those messages. The signature scheme is secure if $\mathcal{A}$ cannot produce,
	 except with some probability negligible in $\lambda$,
	 a valid signature on a message not queried previously, that is, for
	 any probabilistic polynomial-time adversary $\mathcal{A}^{\text{Sign}_{sk}(\cdot)}$,
	 the following probability
	 \beas
	  \Pr[(m,\sigma)\leftarrow\mathcal{A}^{\text{Sign}_{sk}(\cdot)}(1^\lambda):
	  {m\not\in Q} \wedge \text{Verify}(pk,m,\sigma)=\texttt{accept}]
	 \eeas
	 is negligible in $\lambda$, where $Q$ is the set of sign queries made by
	 $\mathcal{A}$ to $\text{Sign}_{sk}(\cdot)$.
\end{enumerate}

As a concrete example, we mention the algorithms of the BLS signature proposed by Boneh,
Lynn and Shacham~\cite{BLS_JOC}. Let the algorithm BLSetup$(1^\lambda)$ output
$(p,g,G,G_T,e)$ as the parameters of a bilinear map, where $G$ and $G_T$ are
multiplicative cyclic groups of prime order $p$, $g$ is a generator of $G$ and
$e:G\times G\rightarrow G_T$ (see Section~\ref{blmap}). KeyGen chooses
$sk\xleftarrow{R} \Z_p$ as the secret key, and the public key is set to be
$pk=g^{sk}$. The algorithm Sign uses a full-domain hash $H:\{0,1\}^*\rightarrow G$,
and it generates a signature $\sigma=H(m)^{sk}$ on a message $m\in\{0,1\}^*$.
Given a message-signature pair $(m,\sigma)$, the algorithm Verify checks
$e(\sigma,g)\stackrel{?}=e(H(m),pk)$. Verify outputs \texttt{accept} if and only if
the equality holds.

\subsection{Erasure Code}\label{erasure_code}
An $(n,f,d)_\Sigma$ erasure code is a forward error-correcting code~\cite{MWSloane77}
that consists of an encoding algorithm Enc: $\Sigma^f\rightarrow\Sigma^n$
(encodes a message consisting of $f$ symbols into a longer codeword consisting of $n$ symbols) and
a decoding algorithm Dec: $\Sigma^n\rightarrow\Sigma^f$ (decodes a codeword to a message),
where $\Sigma$ is a finite alphabet and $d$ is the minimum distance
(Hamming distance between any two codewords is at least $d$) of the code.
The quantity $\frac{f}{n}$ is called the rate of the code.
An $(n,f,d)_\Sigma$ erasure code can tolerate up to $d-1$ erasures. If $d=n-f+1$, we call the code
a maximum distance separable (MDS) code. For an MDS code, the original message can be reconstructed
from any $f$ out of $n$ symbols of the codeword~\cite{Mitzen_ITW}. Reed-Solomon codes~\cite{RSCode}
and their extensions are examples of non-trivial MDS codes. We give a simple example,
from~\cite{MWSloane77}, of a Reed-Solomon code below.

Let us consider the finite field $\F_{2^2}=\{0,1,\alpha,\gamma=\alpha^2\}$ with $\alpha^2+\alpha+1=0$.
A $(3,2,2)$ Reed-Solomon code over $\F_{2^2}$ consists of 16 codewords:
\[\begin{array}{ccccccc}
000 & {\hphantom{000}} & 1\alpha 0 & {\hphantom{000}} & \gamma 0\alpha & {\hphantom{000}} & \gamma\alpha 1\\
01\alpha & {\hphantom{000}} & \alpha\gamma 0 & {\hphantom{000}} & 10\gamma & {\hphantom{000}} & 111\\
0\alpha\gamma & {\hphantom{000}} & \gamma 10 & {\hphantom{000}} & 1\gamma\alpha & {\hphantom{000}} & \alpha\alpha\alpha\\
0\gamma 1 & {\hphantom{000}} & \alpha 01 & {\hphantom{000}} & \alpha 1\gamma & {\hphantom{000}} & \gamma\gamma\gamma.\\\
\end{array}\]
This code can correct a single erasure ($d-1=1$).
For example, $1*\gamma$ (`$*$' denotes the erasure) can be decoded uniquely to $10\gamma$. In other
words, a partially erased codeword can be reconstructed from the other two symbols available.

\subsection{Oblivious RAM}\label{oram}
Goldreich and Ostrovsky introduce the notion of oblivious RAM (ORAM)~\cite{ORAM_JACM}.
In a RAM (Random Access Memory) model, there is a CPU and a memory module. Anyone can
intercept the communications between the CPU and the memory module, and observe the
memory-access patterns. Oblivious RAM (ORAM) is a probabilistic RAM where the access-pattern
is independent of the address input to the memory module.

ORAM involves a hierarchical data structure which allows hiding memory-access patterns.
This data structure consists of hash tables of different lengths at different levels.
The number of levels is $O(\log n)$.
An element of a hash table contains an (\textit{address, value}) pair. When an address
is searched for a read operation, the address is first hashed and the hash value
is matched with the hash table at the top level. If a match is not found, the address is
hashed again and matched with the hash table in the next level, and so on. If a match is
found, random locations are searched in the hash tables in the subsequent levels.
This is continued until the last level. If an address is found more than once, ORAM returns
the most updated value residing at the topmost level. For a write operation, the new value is inserted
into the hash table of the top level. As each address-search is associated with hash tables
in every level of the hierarchical data structure, an adversary cannot gain any knowledge
about the pattern of the search. For the same reason, ORAM takes time polynomial in $\log n$
for each read or write operation as all the hash tables need to be consulted to hide the actual access-pattern.
There is a ``rebuild'' phase which is executed periodically to rebuild the levels (due to too many insertions).
Recent works on ORAM include
\cite{Pinkas_ORAM,Goodrich_ORAM,Stefanov_ORAM1,Stefanov_ORAM2,Stefanov_ORAM3,Stefanov_ORAM4,Shi_ORAM1}.

\subsection{Proofs of Retrievability}\label{por}
A client uploads a file to the cloud server. However, the client needs a
guarantee that \textit{all} her data are stored in the server untampered.
Proofs-of-retrievability (POR) schemes make the client be assured that her data
are stored intact in the server. Juels and Kaliski introduce proofs of retrievability
for static data~\cite{JK_CCS}. Static data mostly include archival
data which the client does not modify after she uploads the file to the server.
However, some of the POR schemes deal with dynamic data where the client modifies her data.
We provide a brief idea about the building blocks of POR schemes. We discuss
them in detail in Section~\ref{por_static} and Section~\ref{por_dynamic}.

\begin{figure}[htbp]
\centering
\includegraphics[width=0.57\textwidth]{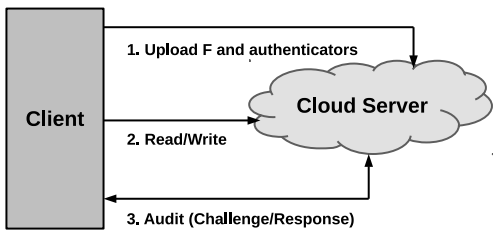}
\caption{Basics of a proofs-of-retrievability scheme.}\label{fig:basics}
\end{figure}

In the setup phase, the client preprocesses her file $F_0$. The preprocessing step involves
encoding the file $F_0$ with an erasure code to form another file $F$. Then, an authenticator
is attached to each of the blocks of $F$ (for checking the integrity of the blocks later).
Finally, the client uploads $F$ along with the authenticators to the server.
We consider the file $F$ as a collection of $n$ blocks or segments where each block
is an element of $\Z_p$. The client can read data from the file
she has outsourced. She performs audits to check the integrity of her data.
An audit comprises of two algorithms for proof-generation and proof-verification. 
During an audit, the client generates a random challenge and sends it
to the server which acts as a prover. Upon receiving the challenge, the server
responds to the client with a proof. The client then verifies the integrity
of the data by checking the validity of the proof. If the proof is valid, the
verification algorithm outputs 1; otherwise, it outputs 0. For dynamic POR schemes,
the client can issue write operations along with read operations. The basic
operations are illustrated in Fig.~\ref{fig:basics}.

POR schemes satisfy two properties: correctness and soundness. The correctness property
demands that the proof generated by an honest server always makes the verification
algorithm output 1. The soundness property of POR schemes is formalized by the
existence of an extractor algorithm that extracts $F$ after interacting with a malicious server which
passes an audit (that is, the verification algorithm outputs 1) with any probability
non-negligible in the security parameter $\lambda$.

There are two types of POR schemes: \textit{privately verifiable} and \textit{publicly verifiable}
schemes. In private-verification schemes, only the client can perform audits as the verification
of a proof requires some secret information. On the other hand, in publicly verifiable
schemes, anyone can verify the proof supplied by the server. In privacy preserving auditing,
the verifier (any verifier other than the client) cannot gain any knowledge about the data
outsourced to the server~\cite{Wang_TC}.

\section{Proofs of Retrievability for Static Data}\label{por_static}
In the static setting, the client does not modify her data once they
are outsourced to the cloud server. We discuss two POR schemes for
static data below. However, there are other POR schemes related to
static data. We mention some of them in Section~\ref{conclusion}.

\subsection{POR Scheme by Juels and Kaliski}
Juels and Kaliski~\cite{JK_CCS} propose the first POR scheme for static
data. A similar scheme for online memory checking is given by Naor
and Rothblum~\cite{NR_JACM}. Though the basic idea is the same for both
of the schemes, the first one uses a \textit{sentinels} (random strings
that are independent of the file's content) and the latter scheme uses
MACs for authentication. Here, we describe the MAC-based solution and
make a brief note about the sentinel-based solution.

The client selects $k\xleftarrow{R} \mathcal{K}$ as her secret key, where
$\mathcal{K}$ is the key space for a MAC.
Let the client have a file $F_0$ with $f$ blocks or segments which she wants
to upload to the cloud server. The client encodes $F_0$ with an
erasure code to form a file $F$ with $n$ segments.
Let each segment of the file $F$ be an element of $\Z_p$,
that is, $F[i]\in \Z_p$ for all $1\le i\le n$.
The client computes $\sigma_i=\text{MAC}_k(i||F[i])$ for all $1\le i\le n$
and uploads the file $F$ along with the tags $\{\sigma_i\}_{1\le i\le n}$ to the
server.

During an audit, the client generates a random challenge $Q=\{i\}$ and sends it
to the server which acts as a prover. Upon receiving $Q$, the prover
responds to the client with $\{(F[i],\sigma_i)\}_{i\in Q}$. The verification algorithm,
for each $i\in Q$, checks if
 \begin{equation*}
  \sigma_i\stackrel{?}=\text{MAC}_k(i||F[i]),
 \end{equation*}
and outputs 1 if the equality holds for each $i\in Q$; it outputs 0, otherwise.
If the MAC scheme is secure, then the server cannot produce a valid MAC on a
message of its choice without knowing the secret key $k$. Now, the client can
get a fraction (say, $\rho$) of the data blocks of $F$ by interacting with a server which passes
an audit with some probability, non-negligible in the security parameter
$\lambda$. Since the initial file $F_0$ has been encoded to form $F$,
all the blocks of $F_0$ can be retrieved from $\rho$-fraction of blocks
of $F$.

The scheme mentioned above is privately verifiable as only the client
(having the knowledge of the secret key $k$) can verify the integrity of her
data. However, this scheme can be turned into a publicly verifiable scheme
if MACs are replaced by digital signatures.

In the original scheme proposed by Juels and Kaliski~\cite{JK_CCS}, the blocks
of the encoded file $F$ are encrypted and a large number of random elements
(sentinels) are inserted in random locations of $F$. The server cannot distinguish
between the encrypted blocks of $F$ and the sentinels. During an audit, the verifier
(only the client can be the verifier) checks the authenticity of several sentinels
at different positions. If the server modifies a considerable fraction of the blocks,
a similar fraction of sentinels are modified as well (as the sentinels are inserted
in random locations of $F$). The server cannot selectively delete non-sentinel blocks
as it cannot distinguish them from sentinels. Thus, with high probability, the server
cannot pass the audit. On the other hand, once the client challenges for some
sentinel-locations, they are revealed to the server. Therefore,
the future challenges must not include these locations. This makes the number of
audits that can be performed in this scheme bounded.

\subsection{POR Schemes by Shacham and Waters}
\label{cpor}
Shacham and Waters propose two short and efficient homomorphic authenticators
in their POR schemes for static data~\cite{SW_ACR,SW_JOC}. The first one,
based on pseudorandom functions, provides a POR scheme which is privately
verifiable (that is, only the client can verify a proof) and secure in the
standard model\footnote{Standard model is a model of computation where the
security of a cryptographic scheme is derived from some complexity assumptions
(for example, hardness of factoring large integers~\cite{IntFac-Wiki}, or hardness
of finding discrete logarithm of an element of a finite group~\cite{DisLog-Wiki}.)};
the second one, based on BLS signatures (see Section~\ref{dig_sig}), gives a POR
scheme which is publicly verifiable (that is, anyone can verify a proof) and
secure in the random oracle model\footnote{Random oracle model is a model of
computation where the security of a cryptographic scheme is proven assuming
a cryptographic hash function used in the scheme as a truly random function.}
\cite{BR_RO}.

As mentioned by Shacham and Waters, Reed-Solomon codes are necessary
against adversarial erasures where the server can delete
blocks selectively. One drawback of these codes is the complexity of encoding and decoding
is $O(n^2)$, where $n$ is the number of blocks of the file uploaded to the server. We can
employ codes with linear decoding time instead of Reed-Solomon codes. However, these codes
are secure against random erasures only. Shacham and Waters discuss a solution to this
problem strictly for the privately verifiable scheme.
We briefly describe the schemes below.

\subsubsection{POR Scheme with Private Verification}
The client chooses $(\alpha,k)$ as her secret key, where $\alpha\xleftarrow{R} \Z_p$
and $k\xleftarrow{R} \mathcal{K}$ ($\mathcal{K}$ is the key space for a pseudorandom
function). Let $h: \mathcal{K} \times \{0,1\}^*\rightarrow\Z_p$ be a pseudorandom
function~\cite{KatzLindell,GoldreichI,Luby_PRP}.
Let the client have a file $F_0$ with $f$ blocks or segments which she wants
to upload to the cloud server. The client encodes $F_0$ with an
erasure code to form a file $F$ with $n$ segments.
Let each segment of the file $F$ be an element of $\Z_p$,
that is, $F[i]\in \Z_p$ for all $1\le i\le n$.
The client computes $\sigma_i=h_k(i)+\alpha F[i] \mod p$ for all $1\le i\le n$
and uploads the file $F$ along with the tags $\{\sigma_i\}_{1\le i\le n}$ to the
server.

During an audit, the client generates a random query $Q=\{(i,\nu_i)\}$ and sends it
to the server which acts as a prover. Upon receiving $Q$, the prover computes
$\sigma=\sum_{(i,\nu_i)\in Q}{\nu_i\sigma_{i}}\mod p$ and $\mu=\sum_{(i,\nu_i)\in Q}\nu_iF[i]\mod p$.
The prover responds to the client with $(\sigma,\mu)$. Then the client verifies
the integrity of her data by checking the verification equation
 \begin{equation*}
  \sigma\stackrel{?}=\left(\alpha\mu+\sum_{(i,\nu_i)\in Q}{\nu_i h_k(i)}\right)\mod p,
 \end{equation*}
and outputs 1 or 0 depending on whether the equality holds or not. As discussed in
Section~\ref{por}, a POR scheme is correct if the verifier always outputs 1 when the
proof is supplied by an honest server. The correctness of the scheme can be proved
as below.
\begin{equation*} 
 \begin{split}
  \text{Correctness:}\qquad\sigma 	& \cong \sum\limits_{(i,\nu_i)\in Q}{\nu_i\sigma_{i}}\\
		& \cong \sum\limits_{(i,\nu_i)\in Q}{\nu_i\left(h_k(i)+\alpha F[i]\right)}\\
		& \cong \alpha\sum\limits_{(i,\nu_i)\in Q}{\nu_i F[i]} + \sum\limits_{(i,\nu_i)\in Q}{\nu_i h_k(i)}\\
		& \cong \left(\alpha\mu + \sum\limits_{(i,\nu_i)\in Q}{\nu_i h_k(i)}\right)\mod p
 \end{split}
\end{equation*}
In this privately verifiable scheme, only the client can perform the verification as
the verification algorithm requires the knowledge of the secret key $(\alpha,k)$.

\subsubsection{POR Scheme with Public Verification}
Let there be an algorithm BLSetup$(1^\lambda)$ that outputs $(p,g,G,G_T,e)$
as the parameters of a bilinear map, where $G$ and $G_T$ are multiplicative
cyclic groups of prime order $p=\Theta(\lambda)$, $g$ is a generator of $G$ and
$e:G\times G\rightarrow G_T$ (see Section~\ref{blmap}). The client chooses
$x\xleftarrow{R} \Z_p$ as her secret key. The public key of the client is
$v=g^x$. Let $\alpha\xleftarrow{R} G$ be another generator of $G$ and
$H:\{0,1\}^*\rightarrow G$ be the BLS hash (see Section~\ref{dig_sig}).
Let the client have a file $F_0$ with $f$ blocks or segments which she wants
to upload to the cloud server. The client encodes $F_0$ with an
erasure code to form a file $F$ with $n$ segments.
Let each segment of the file $F$ be an element of $\Z_p$,
that is, $F[i]\in \Z_p$ for all $1\le i\le n$.
The client computes $\sigma_i=(H(i)\cdot{\alpha}^{F[i]})^x$ for all $1\le i\le n$
and uploads the file $F$ along with the tags $\{\sigma_i\}_{1\le i\le n}$ to the
server.

During an audit, the verifier generates a random query $Q=\{(i,\nu_i)\}$ and sends it
to the server which acts as a prover. Upon receiving $Q$, the prover computes
$\sigma=\prod_{(i,\nu_i)\in Q}{\sigma_{i}}^{\nu_i}$ and $\mu=\sum_{(i,\nu_i)\in Q}\nu_iF[i]\mod p$.
The prover responds to the verifier with $(\sigma,\mu)$. Then the verifier verifies
the integrity of client's data by checking the verification equation
 \begin{equation*}
  e(\sigma,g)\stackrel{?}=e\left(\prod_{(i,\nu_i)\in Q}H(i)^{\nu_i}\cdot{\alpha}^{\mu},v\right),
 \end{equation*}
and outputs 1 or 0 depending on whether the equality holds or not. The correctness
of the scheme can be proved as below.
\begin{equation*} 
 \begin{split}
  \text{Correctness:}\qquad\sigma 	& = \prod\limits_{(i,\nu_i)\in Q}\sigma_{i}^{\nu_i}\\
		& = \prod\limits_{(i,\nu_i)\in Q}(H(i)\cdot\alpha^{F[i]})^{\nu_ix}\\
		& = \left(\prod\limits_{(i,\nu_i)\in Q}H(i)^{\nu_i}\cdot\prod\limits_{(i,\nu_i)\in Q}\alpha^{\nu_i F[i]}\right)^{x}\\
		& = \left(\prod\limits_{(i,\nu_i)\in Q}H(i)^{\nu_i}\cdot\alpha^{\sum_{(i,\nu_i)\in Q}\nu_i F[i]}\right)^{x}\\
		& = \left(\prod\limits_{(i,\nu_i)\in Q}H(i)^{\nu_i}\cdot\alpha^\mu\right)^{x}
 \end{split}
\end{equation*}
In this publicly verifiable scheme, the verifier does not need the secret key $x$ to
verify the response from the prover; knowledge of the public key $pk$ would suffice
for that purpose. Due to this reason, any third party auditor (TPA) can perform audits
on behalf of the client (owner of the data). In privacy preserving auditing, there is
an additional requirement that the TPA should not learn the data on which the audits
are being performed. For example, Wang et al.~use the publicly verifiable
scheme of Shacham and Waters, and they achieve privacy preserving auditing using a
technique called random masking~\cite{Wang_TC}.

\section{Proofs of Retrievability for Dynamic Data}\label{por_dynamic}
In the previous section, we have described some POR schemes for
static data which the clients do not modify once they are uploaded in
the cloud server. A natural question comes if any POR schemes are available
for dynamic data where the clients modify their outsourced data ``efficiently''.
In this section, we discuss about the difficulties of modification of the
uploaded data. Then, we will mention two POR schemes for dynamic data.

To maintain the retrievability of the whole file, erasure coding has been
employed on the file. The blocks of the file are encoded in such a way that
the file can be retrieved from a fraction of blocks of the encoded file.
The content of each block is now distributed in other $O(n)$ blocks.
Therefore, to actually delete a block the server has to delete all the related
blocks. This restricts the server from deleting or modifying a block maliciously
and still passing the verification with non-negligible probability in $\lambda$.
However, this advantage comes with some drawbacks. If the client wants to update
a single block, she has to update all the related blocks as well. This makes the
update process inefficient as $n$ can be very large.

Cash et al.~\cite{Wichs_ORAM} discuss about two failed attempts to provide
a solution of the problem mentioned above. In the first case, a possible solution
might be to encode the file locally. Now, each codeword consists of a small number
of blocks. Therefore, an update of a single block requires an update of a few
blocks within that particular codeword. However, a malicious server can gain the
knowledge of this small set of blocks (within a codeword) whenever the client
updates a single block. Thus, the server can delete this small set of blocks
without being noticed during an audit. In the second attempt, after encoding the
file locally, all of the $n$ blocks are permuted in a pseudorandom fashion.
Apparently, the server cannot get any information about the blocks in a codeword.
However, during an update the server can identify the related blocks in a codeword.
Therefore, the server can again delete these blocks and pass the verification
during an audit.

Due to the issues discussed above, only a few POR schemes for dynamic data are
available in the literature. Now, we briefly mention two of these schemes below.
The first scheme~\cite{Wichs_ORAM} exploits oblivious RAM for hiding data-access
patterns. The second scheme~\cite{Stefanov_CCS} uses an incremental code to reduce
the amortized cost for updates.

\subsection{POR Scheme by Cash, K{\"{u}}p{\c{c}}{\"{u}} and Wichs}

Cash et al.~\cite{Wichs_ORAM} propose a POR scheme for dynamic data using ORAM
(see Section~\ref{oram}). They proceed as the first attempt mentioned in
Section~\ref{por_dynamic}. That is, the data is divided into several chunks
where each chunk contains a few blocks in it. Then, the blocks in each chunk
are encoded ``locally'' using an erasure code to form a codeword. Thus, an
update on a single block requires updating only related blocks of that particular
codeword. This makes the update process much more efficient than that when
all the blocks of the data are encoded to form a single large codeword.
However, this solution comes with a drawback that the malicious server
can now identify all the related blocks and delete these blocks
selectively. As the number of blocks in a codeword is small, the server
has a considerable chance to get through an audit.

Cash et al. introduce ORAM as a solution for the problem mentioned above,
still keeping the update-complexity low. Small chunks are encoded to form
small codewords to make the updates efficient. However, the challenge is
to hide the access-patterns from the server so that the server cannot
identify the blocks in a codeword. ORAM lets the client read from the
outsourced data in a pseudorandom fashion (using ORAM-Read protocol).
It also provides a privacy-preserving way to write the blocks of a codeword
(using ORAM-Write protocol).
We give a high-level overview of the scheme as follows.

In the setup phase, data blocks are divided into chunks and chunks are
encoded to form codewords. In this phase, the ORAM protocol is initiated
as well. For a read operation, the exact location of the block is found
from the chunk-address (\textit{add\_ch}) and the offset (\textit{add\_off}),
and this address is fed into ORAM-Read. For a write operation, \textit{add\_ch}
and \textit{add\_off} are calculated first.
Then, the codeword corresponding to \textit{add\_ch} is obtained (using ORAM-Read)
and decoded. The exact block is located (using \textit{add\_off}) and modified
accordingly. The new chunk is now encoded again and updated in the server using
ORAM-Write. To run the audit protocol, a set of random locations are read using
ORAM-Read and their authenticity is checked. The verifier outputs 1 if and only
if the data-integrity is preserved.

\subsection{POR Scheme by Shi, Stefanov and Papamanthou}

The privacy of the access-patterns is achieved by the scheme proposed by
Cash et al.~\cite{Wichs_ORAM}. However, Shi et al.~\cite{Stefanov_CCS}
argue that a POR scheme need not hide the access-patterns and
it must satisfy only two properties: authenticated
storage (the client needs an assurance about the authenticity of her data)
and retrievability (the client can extract her data at any point of time).
Shi et al.~propose another dynamic POR scheme where the scheme satisfies
these two properties, and it is more efficient (in terms of the computation-cost
or communication bandwidth required for the basic operations) than the scheme
by Cash et al.~as the additional cost for hiding access-patterns is now eliminated.
Here, we describe the basic construction of the scheme briefly.

The main challenge is to reduce the write cost since an update in a single block
is followed by updates on other $O(n)$ blocks. In this scheme, the encoded copy is
not updated for every write operation. Instead, it is updated (or rebuilt) only when
sufficient updates are done on the data file. Thus, the amortized cost for writes is
reduced dramatically. However, between two such rebuilds this encoded copy stores
stale data. Therefore, they introduce a temporary hierarchical log structure which
stores values for these intermediate writes. During an audit, $O(\lambda)$ random
locations of the encoded data file as well as the hierarchical log structure are
checked for authenticity.
The scheme involves three data structures: an uncoded buffer U which is updated
updated after every write and reads are performed on this buffer only, an encoded
(using an erasure code) buffer C which is updated after every $n$ writes, and an
encoded log structure H which accommodates every update between two rebuilds of C.

The buffer U contains an up-to-date copy of the data file. Reads and writes are
performed directly on the required locations of U. Merkle hash tree~\cite{Merkle-Wiki,Merkle_CR}
is used for U to check the authenticity of the read block. Reads and writes are never
performed directly on the buffer C. After $n$ write operations, the buffer U is
encoded using an erasure code (see Section~\ref{erasure_code}) to form a new copy
of C, and the existing copy of C is replaced by this new one.
The log structure H consists of $\log n+1$ levels and stores the intermediate updates
temporarily. The $l$-th level consists of an encoded copy of $2^l$ blocks using
a $(2^{l+1},2^l,2^l)$-erasure code for each $0\le l\le \log n+1$. When a block is
updated it is written in the topmost level ($l=0$). If the top $l$ levels are already
full, a rebuild is performed to accommodate all the blocks up to $l$-th level as well as
the new block in the $(l+1)$-th level and to make all the levels up to $l$-th level empty.
Shi et al.~employs a fast incrementally constructible code based on Fast Fourier Transform~\cite{FFT-Wiki}.
Using this code, the rebuild cost of $l$-th level takes $O(\beta\cdot 2^l)$ time, where
$\beta$ is the block size. The $l$-th level is rebuild after $2^l$ writes. Therefore,
the amortized cost for rebuilding is $O(\beta\log n)$ per write operation. This improves the
earlier scheme of Cash et al.~\cite{Wichs_ORAM} which requires $O(\beta\lambda(\log n)^2)$.
Each rebuild of C is followed by making H empty.
To perform an audit, the verifier chooses $O(\lambda)$ random locations of the encoded buffer C
and $O(\lambda)$ random locations of each full level of the hierarchical log structure H,
and check for authenticity. The verification algorithm outputs 1 if all the blocks are authentic;
it outputs 0, otherwise.

Shi et al.~\cite{Stefanov_CCS} improve this basic construction by using
homomorphic checksums. In this improved construction, the cost of communication
bandwidth and the cost of client computation are further reduced to $\beta+O(\lambda\log n)$.
However, the server computation remains the same, that is, $O(\beta\log n)$.

\section{Conclusion}\label{conclusion}

In this chapter we have given a brief overview of proofs-of-retrievability (POR),
and we have discussed some POR schemes. There are several POR schemes in the
literature which we have not covered in this chapter. Interested readers may take
a look at these works
\cite{Wichs_HA,Stinson_POR,Bowers_CCSW,Bhavana_TCC,Bowers_HAIL,Outpor_CCS,IRIS}.

\end{document}